\documentclass[letterpaper,12pt]{article}   
\usepackage{osajnl2} 
\usepackage[draft]{hyperref} 

\begin{document}

\title{An unconventional geometric phase gate with two non-identical quantum
dots trapped in a photonic crystal cavity}
\author{Jian-Qi Zhang, Ya-Fei Yu, Zhi-Ming Zhang$^{*}$}

\address{Key Laboratory of Photonic Information Technology of Guangdong Higher
Education Institutes, SIPSE $\&$ LQIT, South China Normal
University, Guangzhou 510006, China} \address{$^*$Corresponding
author: zmzhang@scnu.edu.cn}

\begin{abstract}
We propose a scheme for realizing a two-qubit controlled phase gate
via an unconventional geometric phase with two non-identical and
spatially separated quantum dots trapped in a photonic crystal
cavity. In this system, the quantum dots simultaneously interact
with a large detuned cavity mode and strong driving classical light
fields. During the gate operation, the quantum dots undergo no
transitions, while the cavity mode is displaced along a closed path
in the phase space. In this way, the system can acquire different
unconventional geometric phases conditional upon the states of the
quantum dots, and a two-qubit entangling gate can be constructed.
\end{abstract}


\maketitle

\section{Introduction}

As it is well known, a solid state implementation of cavity quantum
electrodynamics (QED)-based approaches would open new opportunities for
scaling the network into the practical and useful quantum information
processing (QIP) systems \cite{01}. Among the proposed schemes, the systems
of self-assembled quantum dots (QDs) embedded in photonic crystal (PC)
nanocavities have been a kind of very promising systems. That is not just
because the strong QD-cavity interaction can be realized in these systems
\cite{11,1101,1102}, but also because both QDs and PC cavities are suitable
for monolithic on-chip integration.

However, there are two main challenges in this kind of systems. One is that
the variation in emission frequencies of the self-assembled QDs is large
\cite{14}. It is very difficult to achieve identical QDs in experiments, so
it is important to realize QIP with non-identical QDs which include both the
near-resonant QDs and the nonresonant QDs. The other is that the interaction
between the QDs is difficult to control \cite{15}. So far, there are several
methods which have been used to bring the emissions of non-identical QDs
into the same, such as, by using Stark shift tuning \cite{16,1601,161} and
voltage tuning \cite{17}. There are also several solutions which have been
used to control the interaction between the QDs, for instance, coherent
manipulating coupled QDs \cite{15}, and controlling the coupled QDs with
Kondo effect \cite{18}. Due to the small line widths of the QDs, cavity
modes and the frequency spread of the QD ensemble, the tuning of individual
QD frequencies is mainly achieved for two closely spaced QDs trapped in a PC
cavity \cite{17}. And the controlled interactions and logical gates between
the QDs can be acquired in this way. Moreover, the entanglement of two QDs
has been achieved in the experiment \cite{171}. On the contrary, there are
few schemes about how to achieve the controlled interactions and logical
gates with the spatially separated QDs \cite{61}.

On the other hand, the quantum gates based on the dynamical phases (DPs) are
sensitive to the quantum fluctuations, which are the main blockage toward a
large-scale quantum computing. The ideas that adopt the geometric phase have
been utilized for solving this problem \cite{19,20,21,22,23,24}. As the
geometric phase is determined only by the path area, it is insensitive to
the starting state distributions, the path shape, and the passage rate to
traverse the close path \cite{20,22}. In this aspect, geometric phases are
better than the dynamical ones in realization of quantum computing. Until
now, there are two kinds of geometric phase gates (GPGs). The gates
containing no DP are often referred to as conventional GPGs \cite{31}. In
contrast, the gates including the DP are named unconventional GPGs \cite{20}%
. Comparing with the conventional GPGs, for the DP in the unconventional
GPGs being proportional to the geometric phase, they don't need to eliminate
the DP. For this reason, the unconventional GPGs are better than the
conventional ones. Moreover, the unconventional GPGs have been realized in
the system of ions \cite{30}.

Recently, Feng \textit{et al.} proposed a scheme to realize a quantum
computation with atoms in decoherence-free subspace by using a dispersive
atom-cavity interaction driven by strong classical laser fields \cite{24}.
But their proposal is based on identical qubits, and each atom is driven
with four laser fields. Motivated by this work, we propose a scheme for
constructing an unconventional GPG with two different QDs trapped in a
single-mode cavity, and each dot is driven with two laser fields. In this
scheme, the QDs undergo no transitions, while the cavity mode is displaced
along a closed path in the phase space. In this way, the system can acquire
different phases conditional upon different states of QDs. With the
application of single-qubit operations, a controlled phase gate can be
constructed in this system. During the gate operation, as the QDs undergo no
transitions, the spontaneous emission of QDs can be ignored. Comparing with
Ref. \cite{24}, the logical gate is extended to non-identical qubits, and
the number of laser fields is decreased. The main differences between our
scheme and previous ones for near-neighbor QD systems \cite{2401,2402} about
the phase gates are that our phase gate is an unconventional geometric gate,
and it can be constructed with the spatially separated QDs.

The organization of this paper is as follows. In Sec.\ref{sec2}, we
introduce the theoretical model and effective Hamiltonian. In Sec.\ref{sec3}%
, we review the definition of the unconventional geometric phase and present
a two-qubit controlled phase gate based on the unconventional geometric
phase. In Sec.\ref{sec4}, we show the simulations and discussions of the
two-qubit operation. The conclusion is given in Sec.\ref{sec5}.

\section{Theoretical Model and effective Hamiltonian}

\label{sec2}

As shown Fig.1, we consider that two charged GaAs/AlGaAs QDs are
non-identical and spatially separated. They are trapped in a single-mode PC
cavity. Each dot has two lower states $|g\rangle =|\uparrow \rangle $, $%
|f\rangle =|\downarrow \rangle $ and two higher states $|e\rangle =|\uparrow
\downarrow \Uparrow \rangle $, $|d\rangle =|\downarrow \uparrow \Downarrow
\rangle $, here ($|\uparrow \rangle $, $|\downarrow \rangle $) and ($%
|\Uparrow \rangle $, $|\Downarrow \rangle $) denote the spin up and spin
down for electron and hole, respectively. At zero magnetic field, the two
lower states are twofold degenerate. The only dipole allowed transitions $%
|g\rangle \leftrightarrow |e\rangle $ and $|f\rangle \leftrightarrow
|d\rangle $ are coupled with $\sigma ^{+}$ and $\sigma ^{-}$ polarization
lights, respectively \cite{17,35}. If the fields in the $\sigma ^{+}$
polarization are applied to our system \cite{fmang}, the transition $%
|g\rangle \leftrightarrow |e\rangle $ in dot $j$ ($=A,B$) is coupled to the
cavity mode with the coupling $g_{j}$ and classical laser fields with the
Rabi frequencies $\Omega _{j}$ and $\Omega _{j}^{^{\prime }}$, while $%
|f\rangle $ and $|d\rangle $ are not affected. The detunings for the cavity
mode and classical fields are $\Delta _{j}^{C}$, $\Delta _{j}$, and $-\Delta
_{j}^{^{\prime }}$, respectively. In our model, the quantum information is
encoded in states $|g\rangle $ and $|f\rangle $. Then the Hamiltonian
describing the interaction between QDs and fields is:

\begin{equation}
\begin{array}{rcl}
\hat{H}_{I} & = & \sum\limits_{j=A,B}(g_{j}ae^{i\Delta _{j}^{C}t}+\frac{%
\Omega _{j}}{2}e^{i\Delta _{j}t}+\frac{\Omega _{j}^{^{\prime }}}{2}%
e^{-i\Delta _{j}^{^{\prime }}t})\sigma _{j}^{+}+H.c.,%
\end{array}
\label{eq1}
\end{equation}%
where $\sigma _{j}^{+}=|e\rangle _{j}\langle g|$, $a$ is the annihilation
operator for the cavity mode.

In order to derive the effective Hamiltonian which can be used to construct
a displacement operator, we use the method proposed in Refs. \cite%
{24,28,2801} and assume the following conditions: (1) $|\Omega _{j}|=|\Omega
_{j}^{^{\prime }}|$; (2) $\Delta _{j}=\Delta _{j}^{^{\prime }}$; (3) the
large detuning condition: $|\Delta _{j}|,|\Delta _{j}^{^{\prime }}|\gg
|g_{j}|,|\Omega _{j}|,|\Omega _{j}^{\prime }|$; (4) $\delta =\Delta
_{j}^{C}-\Delta _{j}$; (5)$|\Omega _{j}|\gg |g_{j}|$. The first condition
together with the second condition can completely cancel the Stark shifts
caused by the classical light fields and related terms. Under the large
detuning condition, if the QDs are initially in the ground states, since the
probability for QDs absorbing photons from the light fields or being excited
is negligible, the excited states will not be populated and can be
adiabatically eliminated. Moreover, the third and the final conditions
ensure that the terms proportional to $|g_{j}|^{2}$ and $|g_{A}g_{B}|$ can
be neglected. The fourth condition can guarantee that $\delta $ is a tunable
constant which is only related to detuning between the cavity field and
light fields. In this situation, the effective Hamiltonian can be written as:

\begin{equation}
\hat{H}_{eff}=-\sum\limits_{j=A,B}(\lambda _{j}ae^{i\delta t}+\lambda
_{j}^{\ast }a^{\dag }e^{-i\delta t})|g\rangle _{j}\langle g|,  \label{eq2}
\end{equation}%
where $\lambda _{j}=\frac{\Omega _{j}^{\ast }g_{j}}{4}(\frac{1}{\Delta
_{j}^{{}}}+\frac{1}{\Delta _{j}^{^{C}}})$. The two terms in Eq.(\ref{eq2})
describe the coupling between the cavity mode and the classical fields
induced by the virtual excited QDs.

As mentioned above, the spatially separated QDs can be addressed
individually and couple independently to their corresponding
classical light fields, we can assume $\lambda _{j}=\epsilon$ which
is achievable by using laser fields with suitable Rabi frequencies.
Then the effective Hamiltonian reduces to:

\begin{equation}
\hat{H}_{eff}=-\sum\limits_{j=A,B}(\epsilon ae^{i\delta t}+\epsilon ^{\ast
}a^{\dag }e^{-i\delta t})|g\rangle _{j}\langle g|.  \label{eq3}
\end{equation}
According to Refs. \cite{20,22,23,24,30}, since the effective Hamiltonian (%
\ref{eq3}) is only related with the cavity mode coupling the two dots, the
time evolution operator takes the form of a displacement operator, it can be
used to realize two-qubit controlled phase gates.

\section{Two qubits controlled phase gate}

\label{sec3}

In this section, we will show how to realize the two-qubit controlled phase
gate with the effective Hamiltonian (\ref{eq3}). At first, we review the
definition of the unconventional geometric phase due to a displacement
operator along an arbitrary path in phase space \cite{20,22,23,24,30}. The
displacement operator is written as follows:

\begin{equation}
D(\alpha )=e^{\alpha a^{\dag }-\alpha ^{\ast }a},  \label{eq0101}
\end{equation}%
where $\alpha (\alpha ^{\ast })$ is a time-dependent parameter, $a^{\dag }$
and $a$ are the creation and annihilation operators of the harmonic
oscillator (which is the mode of the single-mode cavity in this work),
respectively. According to the Baker-Campbell-Hausdorff formula, for a path
consisting of $N$ short straight sections $\Delta \alpha _{m}$, $m=1,2,3,...$%
, the total displacement operation can be expressed as:

\begin{equation}
\begin{array}{rcl}
D_{t} & = & D(\Delta \alpha _{N})...D(\Delta \alpha _{1}) \\
& = & \exp (iIm\{\sum_{m=2}^{N}\Delta \alpha _{m}\sum_{k=1}^{m-1}\Delta
\alpha _{k}^{\ast }\})D(\sum_{m=1}^{N}\Delta \alpha _{m})\mathbf{.}%
\end{array}
\label{eq103}
\end{equation}%
An arbitrary path $\gamma $ can be followed in the limit $N\longrightarrow
\infty $. Thus we can get
\begin{equation}
D_{t}=e^{i\Theta }D(\int_{\gamma }d\alpha ),  \label{eq0104}
\end{equation}%
with $\Theta =Im\{\int_{\gamma }\alpha ^{\ast }d\alpha \}$ being the total
phase which consists of both the geometric phase and the nonzero DP. Since
the nonzero DP is proportional to the geometric phase, the total phase is an
unconventional phase \cite{20}.

When the path is closed, we can obtain
\begin{equation}
D_{t}=D(0)e^{i\Theta }=e^{i\Theta },  \label{eq0106}
\end{equation}
\begin{equation}
\Theta =Im\{\oint \alpha ^{\ast }d\alpha \}.  \label{eq0107}
\end{equation}
In this case, the phase $\Theta $ is determined by the area of a loop in the
phase space and independent of the quantized state of the harmonic
oscillator \cite{22}.

According to the definition of the unconventional geometric phase, in the
infinitesimal interval $[t,t+dt]$, the system governed by the effective
Hamiltonian Eq.(\ref{eq3}) will evolve as follows:%
\begin{equation}
\left\{
\begin{array}{rcl}
|ff\rangle |\theta _{ff}(t)\rangle & \rightarrow & |ff\rangle |\theta
_{ff}(t)\rangle , \\
|fg\rangle |\theta _{fg}(t)\rangle & \rightarrow & e^{-iH_{eff}t}|fg\rangle
|\theta _{fg}(t)\rangle =D(d\alpha _{fg})|fg\rangle |\theta _{fg}(t)\rangle ,
\\
|gf\rangle |\theta _{gf}(t)\rangle & \rightarrow & D(d\alpha
_{gf})|gf\rangle |\theta _{gf}(t)\rangle , \\
|gg\rangle |\theta _{gg}(t)\rangle & \rightarrow & D(d\alpha
_{gg})|gg\rangle |\theta _{gg}(t)\rangle ,%
\end{array}%
\right.  \label{eq101}
\end{equation}
where $d\alpha _{fg}=i\epsilon ^{\ast }e^{-i\delta t}dt$, $d\alpha
_{gf}=d\alpha _{fg}$, $d\alpha _{gg}=2d\alpha _{fg}$, and $|\theta
_{u,v}(t)\rangle (u,v=g,f)$ denotes the state of the cavity mode, which is
determined by the qubit state $|u_{A}\rangle |v_{B}\rangle $\ at the time $t$%
.

If the cavity field is initially in the vacuum state $|0\rangle $, after an
interaction time $t$, the evolution of the system takes the form of

\begin{equation}
\left\{
\begin{array}{rcl}
|ff\rangle |0\rangle & \rightarrow & |ff\rangle |0\rangle , \\
|fg\rangle |0\rangle & \rightarrow & e^{i\phi _{fg}}D(\alpha
_{fg})|fg\rangle |0\rangle , \\
|gf\rangle |0\rangle & \rightarrow & e^{i\phi _{gf}}D(\alpha
_{gf})|gf\rangle |0\rangle , \\
|gg\rangle |0\rangle & \rightarrow & e^{i\phi _{gg}}D(\alpha
_{gg})|gg\rangle |0\rangle ,
\end{array}%
\right.  \label{eq105}
\end{equation}%
with
\begin{equation}
\left\{
\begin{array}{rcl}
\alpha _{fg} & = & i\int\limits_{0}^{t}\epsilon ^{\ast }e^{-i\delta t}dt=-%
\frac{\epsilon ^{\ast }}{\delta }(e^{-i\delta \tau }-1), \\
\alpha _{gf} & = & i\int\limits_{0}^{t}\epsilon ^{\ast }e^{-i\delta t}dt=-%
\frac{\epsilon ^{\ast }}{\delta }(e^{-i\delta \tau }-1), \\
\alpha _{gg} & = & i\int\limits_{0}^{t}2\epsilon ^{\ast }e^{-i\delta
t}dt=\alpha _{gf}+\alpha _{fg},%
\end{array}%
\right.  \label{eq106}
\end{equation}
and
\begin{equation}
\left\{
\begin{array}{rcl}
\phi _{fg} & = & Im(\int\limits\alpha _{fg}^{\ast }d\alpha _{fg})%
\mathbf{=}-\frac{|\epsilon |^{2}}{\delta }(t-\frac{\sin (\delta t)}{\delta }%
), \\
\phi _{gf} & = & Im(\int\limits\alpha _{gf}^{\ast }d\alpha _{gf})=-%
\frac{|\epsilon |^{2}}{\delta }(t-\frac{\sin (\delta t)}{\delta }), \\
\phi _{gg} & = & Im(\int\limits\alpha _{gg}^{\ast }d\alpha
_{gg})=\phi _{gf}+\phi _{fg}+\theta _{gg}%
\end{array}%
\right.  \label{eq107}
\end{equation}

\begin{equation}
\begin{array}{lll}
\theta _{gg} & = & Im(\int \alpha _{gf}^{\ast }d\alpha _{fg}+\int \alpha
_{fg}^{\ast }d\alpha _{gf}) \\
& = & Im(-2\int\limits_{0}^{t}\frac{|\epsilon |^{2}}{\delta }(1-e^{-i\delta
t})idt) \\
& = & -\frac{2|\epsilon |^{2}}{\delta }(t-\frac{\sin (\delta t)}{\delta }).%
\end{array}
\label{eq108}
\end{equation}%
Eq.(\ref{eq106}) shows, under the condition $t=2l\pi /\delta $ , for
$l=1,2,3...$, the displacement parameter $d\alpha _{uv}$ for state
$|u_{A}\rangle |v_{B}\rangle $ moves along a closed path and returns
to the original point in the phase space of the coherent state
$|\alpha _{uv}\rangle $. And the system can acquire the different
unconventional geometric phases (\ref{eq107}) conditional upon the
different states of QDs in this way. Within a definite period of
time, e.g. from $t=2l\pi /\delta $ to $t=2(l+1)\pi /\delta $, the
state for the cavity mode evolves from a vacuum state to a coherence
state at first, and then evolves to the vacuum state again.
Moreover, since Eq.(\ref{eq3}) is only dependent on the detuning
$\delta $ which is not related to the energy-level of QDs, the
two-qubit operation (\ref{eq105}) can be realized with the system
constituted of both non-resonant and near-resonant QDs.

Owing to the detuning $\delta $, after an interaction time $t=2l\pi
/\delta $ , the evolution of the wavefunction for the system
(\ref{eq105}) can be summarized by \cite{30}:
\begin{equation}
\left\{
\begin{array}{rcl}
|ff\rangle |0\rangle & \rightarrow & |ff\rangle |0\rangle , \\
|fg\rangle |0\rangle & \rightarrow & e^{-\Phi}|fg\rangle |0\rangle , \\
|gf\rangle |0\rangle & \rightarrow & e^{-\Phi}|gf\rangle |0\rangle , \\
|gg\rangle |0\rangle & \rightarrow & e^{-4\Phi}|gg\rangle |0\rangle.
\end{array}
\right.  \label{eq109}
\end{equation}
where $\Phi=2l\pi|\epsilon |^{2}/\delta ^{2}$. This two-qubit
operation is induced by the cavity mode which couples the two dots.
In this situation, since the displacement parameters satisfy
$d\alpha _{gg}=2d\alpha _{gf}=2d\alpha _{fg}$, according to
Eqs.(\ref{eq106}) and (\ref{eq107}), the relationship for the
amplifies of the coherent states is $\alpha _{gg}=2\alpha
_{fg}=2\alpha _{gf}$, and the closed area for $\phi _{gg}$ is four
times of the one for $\phi _{fg}=\phi _{gf}$. So unconventional
geometric phases satisfy $\phi _{gg}=4\phi _{fg}=4\phi _{gf}$.

After the application of the single-qubit operations $|g\rangle
_{j}\rightarrow e^{\Phi}|g\rangle _{j}$ \cite{26}, two-qubit
operation (\ref{eq109}) transforms into
\begin{equation}
\left\{
\begin{array}{rcl}
|ff\rangle |0\rangle & \rightarrow & |ff\rangle |0\rangle , \\
|fg\rangle |0\rangle & \rightarrow & |fg\rangle |0\rangle , \\
|gf\rangle |0\rangle & \rightarrow & |gf\rangle |0\rangle , \\
|gg\rangle |0\rangle & \rightarrow & e^{-2\Phi}|gg\rangle |0\rangle .%
\end{array}%
\right.  \label{eq111}
\end{equation}
The above equation represents an unconventional geometric phase gate. In
this quantum phase gate operation, if and only if both qubits are in the
state $|g\rangle $, there will be an additional phase $-2\Phi$ in the system.

\section{Simulations and Discussions}

\label{sec4}

Finally, we present some numerical simulations to demonstrate our
proposal can be realized in current experimental setups and it can
robust against prevailing uncertainties and fluctuations of
parameters to some extend. For the sake of convenience, we take the
two-qubit operation (\ref{eq109}) with $\Phi=\pi/2$ as an example to
discuss the realization. Under the condition of large detuning, the
influences of spontaneous emission from the excited states of QDs
can be ignored, and the main decoherence effect in this system is
due to cavity decay. Then the master equation can be written as
follows:
\begin{equation}
\begin{array}{rcl}
\dot{\rho} & = & -i[H_{I},\rho ]+\frac{\gamma }{2}(2a\rho a^{+}-a^{+}a\rho
-\rho a^{+}a),%
\end{array}
\label{eq02221}
\end{equation}%
where $\rho $\ is the reduced density operator of the system, $\gamma $\ is
the cavity decay rate.

The fidelity of the two-qubit operation can be expressed as:
\begin{equation}
F=\langle \Psi |\rho (T)|\Psi \rangle ,  \label{eq402}
\end{equation}%
where $T=\pi \delta /(2|\epsilon |^{2})$ is the shortest two-qubit
operation time, $|\Psi \rangle $\ is a target state which is
dependent on the initial state. For instance, if the initial state
is in the state $|\Psi _{in}\rangle =\frac{(x|ff\rangle +y|gf\rangle
+z|fg\rangle +w|gg\rangle )}{ \sqrt{x^{2}+y^{2}+z^{2}+w^{2}}}$ with
random coefficients $\{x,y,z,w\}$, the corresponding target state
takes the form of $|\Psi \rangle =\frac{ (x|ff\rangle -iy|gf\rangle
-iz|fg\rangle +w|gg\rangle )}{\sqrt{ x^{2}+y^{2}+z^{2}+w^{2}}}$. The
numerical calculations for the fidelities of the two-qubit operation
versus the cavity decay rate are given in Fig. \ref{fidelity}, and
these fidelities are the average over fidelities for 500 different
initial states $|\Psi _{in}\rangle $. In doing so, some experimental
parameters in Refs.\cite{35,36,3601} are referred.

Fig. \ref{fidelity} shows the following: Firstly, with the increase
of $\gamma /\gamma _{0}$, here $\gamma _{0}=(5ns)^{-1}$ is the
cavity decay rate that has been achieved in the
experiment\cite{36,3601}, the fidelity for the two-qubit operation
decreases. It means that the cavity decay affects the fidelity of
the two-qubit operation largely \cite{24}. The reason for this is
that the state of cavity mode evolves between the vacuum state and
the coherent state. It is worth pointing out that the decay of
coherent state depends on the mean photon number of coherent state
and the cavity decay. On the one hand, when the mean photon number
is definite, the decay of coherent state increases with increasing
the cavity decay. On the other hand, when the cavity decay is
definite, the decay of coherent state increases with increasing the
mean photon number of coherent state. Secondly, with the increase of
$\delta $, the fidelity of the two-qubit operation increases for the
decrease of the mean photon number of coherent state. Thirdly, the
fidelity for $\delta =0.25g_{A}(\delta =2g_{A})$ is about
$99.98\%(99.95\%)$ when the cavity decay rate is $\gamma =\gamma
_{0}$, and it decreases to about $99.96\%(99.88\%)$ when $\gamma
=2\gamma _{0}$. Moreover, as accumulated unconventional geometric
phase for one loop would be small, our system has to take
multi-loops. For this reason, our scheme needs a good cavity, which
can prevent the photons leaking from the cavity and ensure the
higher fidelity. In addition, according to the parameters in Fig. \ref%
{fidelity}, the two-qubit operation time for $\delta =0.25g_{A}(\delta
=2g_{A})$ is about $1.7ns(13.5ns)$ which is much smaller than the effective
decay time of cavity $\gamma /(\frac{|\epsilon |^{2}}{\delta ^{2}})$ $\sim $
$120ns(4000ns)$. Therefore, it is possible to realize our scheme in the
experiment.

Besides the effect of the decoherence, there are some other factors
that affect the fidelity of the two-qubit operation, the prevailing
factors in our scheme are the uncertainties and fluctuations of
parameters, for example, they might add unwanted phases to the
two-qubit operation. According to Ref. \cite{38}, we can calculate
the fidelity of the two-qubit operation versus the parameter
fluctuation as shown in Fig.\ref{para}. For the sake of simplicity,
we assume $\zeta =\Delta k_{j}/k_{j}>0$ with $\Delta k_{j}$ being
the corresponding parameter fluctuation. In this case, both QDs
undergo the same parameter fluctuations when $\zeta$ is definite.
Fig.\ref{para} shows that the variation trends for all curves of the
fidelity are the same. In the absence of parameter fluctuations, the
fidelities of two-qubit operation can be as high as $F=99.95\%$.
When the parameter fluctuations are included, fidelities decrease
with the increase of parameter fluctuations. The fidelity for $\zeta
=0.02$ is about $99.62\%$, and it decreases to about $98.81\%$ when
$\zeta =0.04$. Therefore, to some extent our scheme can also resist
against errors due to the parameter fluctuations. Note that, each
curve in Fig.\ref{para} is only corresponding to one kind of
parameter fluctuation.

\section{Conclusion}

\label{sec5}

In conclusion, we have shown that in a single-mode PC cavity, two
non-identical and spatially separated QDs driven by the classical light
fields can be used to realize the two-qubit controlled phase gate. During
the gate operation, the QDs remain in their ground states, while the cavity
mode is displaced along a circle in the phase space, and the system can
acquire different unconventional geometric phases conditional upon the
states of QDs. After the application of the single-qubit operations, the
controlled phase gate based on the unconventional geometric phase can be
constructed. The distinct advantages of the proposed scheme are the follows:
firstly, as this controlled phase gate is based on the unconventional
geometric phase, it can resist against the parameter uncertainties and
fluctuations to some extent \cite{20,30}; secondly, as the QDs are
non-identical, it is more practical in the experiment; thirdly, as the
quantum information is encoded in the two ground states, this gate is
insensitive to the spontaneous emission of the QDs \cite{22}. Finally, we
have to point out the shortcoming in our scheme is that the single qubit
operations cannot be realized with our model directly.

\section*{Acknowledgments}

The authors thank Prof. Xun-Li Feng for helpful discussions. This work was
supported by the National Natural Science Foundation of China (Grant No.
60978009) and the National Basic Research Program of China (Grant Nos.
2009CB929604 and 2007CB925204).

\clearpage

\section*{List of Figure Captions}

\noindent Fig. 1. (Color online) (a) Schematic diagram of the system. (b)
the configuration of the QDs level structure and relevant transitions. The
states $|g\rangle $ and $|f\rangle $ correspond to two lower levels, while $%
|e\rangle $ and $|d\rangle $ are two higher levels. The transition $%
|g\rangle \leftrightarrow |e\rangle $ for each dot is driven by the cavity
field and the classical pulses with the detunings $\Delta _{j}^{C}$, $\Delta
_{j}$ and $\Delta _{j}^{^{\prime }}$, respectively. $g_{j}$ represents the
coupling rate of the QDs to cavity mode, $\Omega _{j}$ and $\Omega
_{j}^{^{\prime }}$ are the Rabi frequency of the classical pulses.

\noindent Fig. 2. (Color online) Numerical simulation of the fidelity of the
two-qubit operation (\ref{eq109}) versus the cavity decay, with the
parameters $g_{A}=0.10meV$, $g_{B}=0.08meV$, $\Omega _{A}=10meV$, $\Omega
_{B}=13.75meV$, $\gamma _{0}=(5ns)^{-1}$. The detunings of blue line are
given by $\Delta _{A}=200.00meV$, $\Delta _{B}=220.00meV$, and the detunings
of green line are given by $\Delta _{A}=200.09meV$, $\Delta _{B}=220.09meV$,
respectively.

\noindent Fig. 3. (Color online) Numerical simulation of the
fidelity of the two-qubit operation (\ref{eq109}) versus the
parameter fluctuations. The decay rate for cavity is given by
$\gamma=\gamma_{0}$. All other parameters are the same for the blue
line in Fig. \ref{fidelity}.



\clearpage

\begin{figure}[tbph]
\includegraphics[width=8cm]{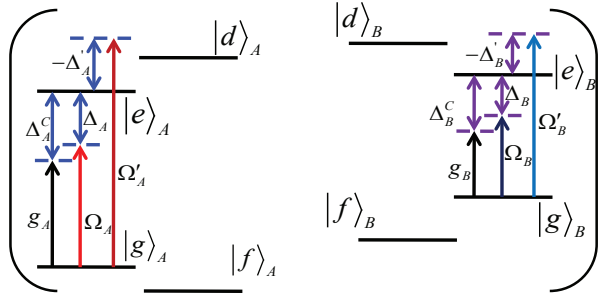}
\caption{(Color online) (a) Schematic diagram of the system. (b) The
configuration of the level structure and relevant transitions for dot $j$ ($%
=A,B$). The states $|g\rangle_j $ and $|f\rangle_j $ correspond to two lower
levels, while $|e\rangle_j $ and $|d\rangle_j $ are two higher levels. The
transition $|g\rangle_j \leftrightarrow |e\rangle_j $ is coupled to the
cavity mode with $g_{j}$ and classical laser fields with the Rabi
frequencies $\Omega _{j}$ and $\Omega _{j}^{^{\prime }}$. The detunings for
the cavity mode and classical fields are $\Delta _{j}^{C}$, $\Delta _{j}$,
and $-\Delta _{j}^{^{\prime }}$, respectively. All the light fields are in
the $\protect\sigma^+$ polarization.}
\end{figure}

\begin{figure}[tbph]
\includegraphics[width=8cm]{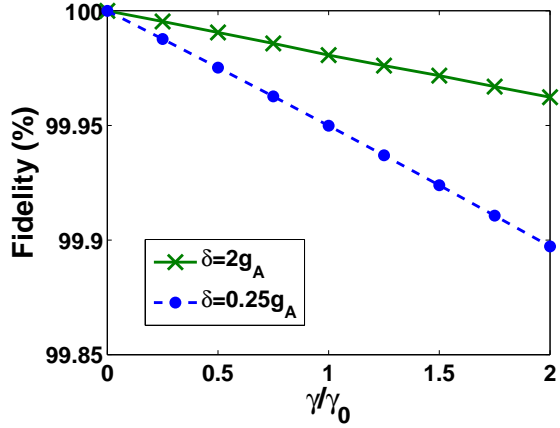}
\caption{(Color online) Numerical simulation of the fidelity of the
two-qubit operation (\protect\ref{eq109}) versus the cavity decay, with the
parameters $g_{A}=0.10meV$, $g_{B}=0.08meV$, $\Omega _{A}=10meV$, $\Omega
_{B}=13.75meV$, $\protect\gamma _{0}=(5ns)^{-1}$. The detunings of blue line
are given by $\Delta _{A}=200.00meV$, $\Delta _{B}=220.00meV$, and the
detunings of green line are given by $\Delta _{A}=200.09meV$, $\Delta
_{B}=220.09meV$, respectively. }
\label{fidelity}
\end{figure}

\begin{figure}[tbph]
\includegraphics[width=8cm]{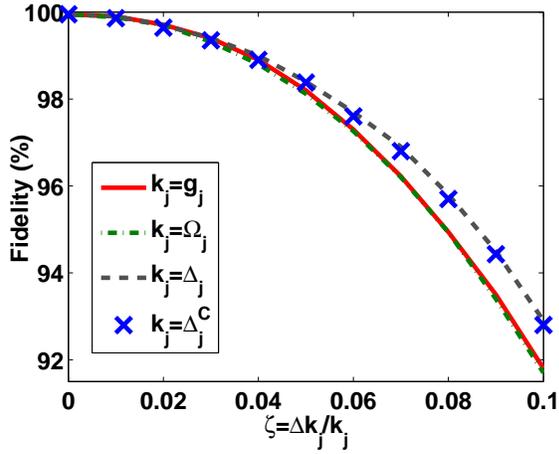}
\caption{(Color online) Numerical simulation of the fidelity of the
two-qubit operation (\protect\ref{eq109}) versus the parameter fluctuations.
The decay rate for cavity is given by $\protect\gamma _{0}=(5ns)^{-1}$. All
other parameters are the same to the blue line in Fig. \protect\ref{fidelity}%
.}
\label{para}
\end{figure}

\end{document}